\documentclass[conference]{IEEEtran}
\usepackage{amssymb,amsmath}
\usepackage{inputenc}
\usepackage{graphicx}
\usepackage{subfig}
\usepackage{algorithm}
\usepackage{algorithmic}
\usepackage{color, soul}
\usepackage{cite}
\usepackage{amsthm}
\usepackage{array}
\sethlcolor{green}
\usepackage{cases}

\definecolor{aaltopurple}{RGB}{102,57,183}  % 6639B7

\newcommand{\Ncopy}{K}
\newcommand{\radius}{r}

\newcommand{\SearchableSet}{\mathcal{S}^{c}}
\newcommand{\ReachableSet}{\mathcal{S}^{r}}

\newcommand{\realPopularity}{\mathbf{p}}

\newcommand{\CouplingCorrelation}{CPF}

\makeatletter
\def\ps@headings{%
\def\@oddhead{\mbox{}\scriptsize\rightmark \hfil \thepage}%
\def\@evenhead{\scriptsize\thepage \hfil \leftmark\mbox{}}%
\def\@oddfoot{}%
\def\@evenfoot{}}
\makeatother
\ifCLASSINFOpdf
\else
\fi

\begin{document}

\title{Effects of Cooperation Policy and Network Topology on Performance of In-Network Caching}

%\title{Coupling Content with Topology:\\ Cooperation Policy in Information-Centric Networks}
% Other optional titles are as follows:
% \title{Towards Pareto Frontier: Cooperation Policy in Information-Centric Network}
% \title{Cooperation Policy: Couple the Content with Topology on System Pareto Frontier}
% \title{On System Pareto Frontier: Couple Content and Topology with Cooperation Policy}
% \title{Effect of Cooperation Policy on Content Placement and Conflicting Performance Metrics in Information-Centric Networks}

\author{\IEEEauthorblockN{Liang Wang, Suzan Bayhan, Jussi Kangasharju}\\
\IEEEauthorblockA{Department of Computer Science, University of Helsinki, Finland}}

\maketitle

\begin{abstract}

In-network caching is a key component in information-centric networking.
In this paper we show that there is a tradeoff between two common caching metrics, byte hit rate and footprint reduction, and show that a cooperation policy can adjust this tradeoff.
We model the cooperation policy with only two parameters -- search radius $r$ and number of copies in the network $\Ncopy$.
These two parameters represent the range of cooperation and tolerance of duplicates.
We show how cooperation policy impacts content distribution, and further illustrate the relation between content popularity and topological properties.
Our work leads many implications on how to take advantage of topological properties in in-network caching strategy design.
 
\end{abstract}

\section{Introduction}
\label{sec:introduction}

%% Background

Caching is a key component in information-centric networks (ICN)~\cite{jacobson:ccn, koponen:dona, psirp, netinf}.
In-network caching not only reduces an ISP's outgoing traffic, but also reduces traffic within an ISP network.
Byte hit rate (BHR) is a common metric for evaluating savings in inter-ISP traffic, however, there is no widely accepted metric for evaluating savings in intra-ISP traffic.
Footprint reduction (FPR)~\cite{anand:smartre} has been proposed as one such metric and is the one we use in this paper.
We show that BHR by itself is insufficient in capturing the performance of a network of caches; this is often overlooked by existing work.
This paper shows that there is a subtle interplay between BHR and FPR and that in some cases these two metrics oppose each other.
% While byte hit rate (BHR) is widely used as an important performance metric to measure the savings on inter-ISP traffic, footprint reduction (FPR) which measures intra-ISP traffic is unfortunately overlooked by the ICN community as another crucial performance indicator.
% In this work, we show that BHR alone may fall short of capturing the ICN performance and sometimes conflicts with the FPR.
We argue that a \emph{cooperation policy} among routers can mediate this tradeoff between BHR and FPR and show that two parameters can tune the desired operating region: maximum number of duplicates for each content item ($\Ncopy$) and the radius for cooperation ($\radius$).

To improve BHR, a cooperation policy covering a large radius enhances network storage utilization by reducing the number of duplicates (\textit{cache diversity} in~\cite{rossini:EvaluatingCCNMultipath_ComCom2013}).
However, large-scale cooperation causes communication overheads and increases intra-ISP traffic because requests may be redirected many times.
Despite efforts in designing a cooperation policy~\cite{licoordinating,6522566,6566743}, a proper model for its impact on BHR and FPR is still missing.
We characterize cooperation policy by its search strength ($\radius$) and capability of reducing duplicates ($\Ncopy$).
We show how different $\radius$ and $\Ncopy$ values lead to different tradeoffs between BHR and FPR, and discuss their implication.% on cooperation policies.

% The proposed model shows different preference on BHR and FPR requires different $\radius$ and $\Ncopy$, further leads to different cooperation policy designs.

There is considerable interest in exploiting topological properties in cache networks.
Initial efforts~\cite{6193506,Chai:2012:CLM} indicate centrality as a promising metric, but questions like how to measure the topological impact on performance and mechanism of the interplay between topology, and caching strategy still remain open.
%Since cooperation policy focuses on the question ``where to cache'', it naturally embraces topology-awareness feature.
We use a cache cooperation policy to couple content with topology and show that this coupling explains how topological properties impact caching performance; the tightness of the coupling indicates degree of topology's impact.

% \textbf{THE FOLLOWING NEEDS MORE CLARITY}
% We propose that as cooperation policy couples content with topology which can be used to explain the mechanism of how topological properties impact performance.
% The tightness of coupling indicates the significance of topological impacts.

%% Contribution

Our contributions in this paper are as follows:

\begin{enumerate}

\item We highlight the importance of FPR as a performance metric for in-network caching, and show how BHR and FPR conflict each other at the Pareto frontier.

\item We propose a cooperation policy model to show the relationship between cooperation policy, content, and topology. We also categorize different cooperation types.

\item We propose a novel way to measure the impact of topology, and perform a thorough numerical analysis to show how it influences system performance.

\end{enumerate}

% In-network caching aims at reducing both intra- and inter-ISP traffic, our work shows these two objectives become conflicting when system reaches its Pareto frontier. Our work also shows simple en-route caching leads to under-utilized system and cooperation policy pushes the performance to its Pareto frontier. As far as we know, we are the first to model the cooperation policy by using only two parameters - search radius and ncopy.

\section{System Model}
\label{sec:sysmodel}

Consider a network of $M$ routers, $L$ of which directly receive user requests and are \textit{edge routers}.
A router denoted by $R_i$ is equipped with a storage capacity of $C_{i}$ bytes.
We assume $N$ distinct files, denoted by $f_i$ and being $s_i$ bytes in size.
All files are stored permanently at the Content Provider (CP) represented as the $(M{+}1)^\mathrm{th}$ router ($R_{M{+}1}$).
%Frequency of request for a file $f_i$ may differ from the others depending on its popularity.
%
Denote the request probability of $f_i$ by this file's \textit{popularity} $p_i$, and denote the popularity vector by $\realPopularity=[p_i]$.
When a request for $f_i$ arrives to an edge router $R_j$, $R_j$ first searches $f_i$ in its cache.
If $R_j$ possesses it, $R_j$ transmits $f_i$ to the user; this a \textit{hit}.
Otherwise, in case of a \textit{miss}, $R_j$ contacts routers in its \emph{r-hop neighborhood} to see if any of them has $f_i$; this is the \emph{cooperation policy}.
%due to the implemented \textit{cooperation policy} between the routers, $R_k$ accesses to the list of items that are stored in the caches of routers in its \textit{r-hop neighborhood}.
We call the set of all routers located at most $r$-hops away from $R_j$ as the \textit{searchable set} of $R_j$, denoted by $\SearchableSet_j$.
If $f_i$ is stored in $\SearchableSet_j$, it is retrieved to $R_j$ from the closest router (if multiple routers holding $f_i$) and forwarded to the user. Let $\mathcal{R}_{j,CP}$ be the set of all routers on the path between a leaf router
   $R_j$ and the CP (excluding the CP). If no router in $\SearchableSet_j$ stores the item, the request is routed to the next router in $\mathcal{R}_{j,CP}$ and searched there as well as in the new searchable set; there may be overlap between searchable sets of two neighboring routers, depending on $r$.
We define the \textit{reachable set} of a router denoted by $\ReachableSet_j$ as the set of all routers in the searchable sets of routers in $\mathcal{R}_{j,CP}$. If no router in $\ReachableSet_j$ has $f_i$, it is downloaded from the CP and routed to the user following the backward path.

\section{Optimal In-Network Caching}
% alternative section heading 
%\section{Optimal In-Network Caching}
\label{sec:optmodel}

\subsection{Cooperation Policy Design}
\label{sec:coop-policy-design}

Performance of a cooperation policy is determined by contents in the searchable set which is a function of $\radius$.
The diversity of contents cached in this set increases caching efficiency which then calls for a caching scheme avoiding duplicate copies in the set~\cite{Wang:2013:ICN, 6522566}.
However, popular content may better be cached in multiple routers to be more accessible from all network edge routers.
We model this tradeoff with parameter $\Ncopy$ which is the maximum number of content replicas in the network.
In reality every file would have its own maximum number of copies which emerges automatically if $\radius$ is fixed; we use a fixed $\Ncopy$ to illustrate system behavior across the whole parameter range.
Using these two parameters, we name a cooperation policy with parameters $K$ and $r$ as \emph{(K, r)-Cooperation Policy} which can be classified into four as follows: 

\begin{enumerate} 
\item Type I, \textbf{small $\radius$, small $\Ncopy$}: Weak cooperation due to limited access to other caches and limited availability of popular content; the system is not using all its resources.
%Both BHR and FPR can be improved, caching system is under-utilized.
\item Type II, \textbf{small $\radius$, large $\Ncopy$}: This is en-route caching. The most popular content is pushed to the network edge.
\item Type III, \textbf{large $\radius$, small $\Ncopy$}: Network storage is effectively a single cache. Popular content is in network core.
\item Type IV, \textbf{large $\radius$, large $\Ncopy$}: Strong cooperation. BHR and FPR cannot be improved at the same time since caching system is fully-utilized and reaches its Pareto frontier.
\end{enumerate}

The complexity of cooperation can be calculated via communication and computation overhead~\cite{6522566}.
Initially, all routers exchange their set of stored contents with routers in their searchable set.
Assuming that each content is unit size and dropping the router index, this initialization step requires $O(MC|\SearchableSet|)$ messages and results in $O(M^2C)$ message exchanges in the worst case.
%For a regular tree topology with $m$ branches and $r$-hop search radius, $\SearchableSet$ has maximum $O(m^r)$ elements.
%In the worst case, this step requires $O(M^2C)$ message exchanges. 
 Upon a change in the cache of a router, this router informs all its $r$-hop neighbours about the evicted and admitted items. This per change announcement requires $O(|\SearchableSet|)$ message in the worst case.
In terms of computation, the cooperation does not involve any processing rather than discovering which of the replicas is closest to a specific router.
Therefore, computation overhead is $O(|\SearchableSet|)$.

\subsection{Optimal Caching under (K, r)-Cooperation Policy} \label{sec:optim-cach-under}

Assume a centralized entity deciding which items are stored at each router $R_i$ when a user requests item $f_u$ at time $t$.
This entity knows the content distribution $\textbf{X}^t = [x^t_{i,j}]$ where $x^t_{i,j}$ is 1 if $f_{i}$ is stored at node $R_{j}$, and zero otherwise.
An optimal caching strategy ($C_\mathrm{OPT}$) determines whether to cache $f_u$ in the routers between the edge router $R_l$ receiving the request and router $R_{hit}$ storing $f_u$, and which items to evict in case of full cache occupancy.
We refer the set of all these intermediate nodes on the path between $R_l$ and $R_{hit}$ as $\mathcal{S}$.

$C_\mathrm{OPT}$ minimizes the total cost of serving user requests by exploiting its knowledge of current content distribution $\textbf{X}^t$, file popularities ($\bf p$), and file size ($s_{i}$) information.
Let $c_{j,k}$ denote the cost of downloading one byte at $R_j$ from $R_k$.
An item can be served from edge router $R_j$ or retrieved from another router $R_{k}$ including the CP.
Let our decision variable $x^{t+1}_{i,j,k}$ be 1 if $R_j$ downloads $f_{i}$ from $R_{k}$.
The cost function reflects the distance between the two entities and can be calculated using shortest path algorithms.  For a (K, r)-Cooperation Policy, as the routers not in $\ReachableSet_j$ are not reachable from this edge router, we set $c_{j,k}=\infty$ if $R_k\not\in\ReachableSet_j$.
% Recall that an item can be served from the edge router $R_j$ itself or it can be retrieved from another router $R_{k}$ including the CP. 
For harmony of notation, we re-define the content distribution by $\textbf{X}^t=[x^t_{i,j,j}]$ (and drop $t$ if we do not refer to a specific time).
$C_\mathrm{OPT}$ is formulated as:
\begin{align}
 &\min \left(\sum_{i=1}^N\sum_{j=1}^L\sum_{k=1}^{M{+}1}s_{i}p_{i}c_{j,k}x^{t{+}1}_{i,j,k}x^{t{+}1}_{i,k,k}
  {+} s_{u}p_u\sum_{j=1}^L\sum_{\begin{subarray}{c}\forall R_{k}\in\\
      \mathcal{S}{\cup}R_{M{+}1} \end{subarray}}c_{j,k}x^{t{+}1}_{u,j,k} \right)
  \label{obj:MinCost}\nonumber\\
  &s.t. \text{ Cache capacity constraints:} \\ &\sum_{i=1}^Ns_{i}x_{i,j,j}^{t}x^{t{+}1}_{i,j,j}{+}s_{u}x^{t{+}1}_{u,j,j}(1{-}x^{t}_{u,j,j}) {\leq} C_j, \forall R_j\in  \mathcal{S}\label{const:CacheCapacity}\\
    &\sum_{i=1}^Ns_{i}x^{t{+}1}_{i,j,j}{\leq} C_j, \forall R_j\not\in  \mathcal{S} \label{const:CacheCapacityNotInS}\\
  &\text{Maximum replica constraint:} \sum_{j=1}^M x^{t+1}_{i,j,j}  \leq \Ncopy \quad \forall i, \label{const:Ncopy}\\
&\text{Feasibility constraints: }  x^{t+1}_{i,j,k} \leq x^{t+1}_{i,k,k} \quad \quad \forall i, \forall k \label{const:FetchFromOther}\\
   &\qquad \qquad \qquad \qquad \quad x^{t+1}_{i,j,j}= x^{t}_{i,j,j} \quad \forall i, \forall R_j \not\in \mathcal{S} \label{const:NoChangeInContents}\\  
   &\text{Service constraint: } 1 \leq \sum_{k=1}^{M+1} x^{t+1}_{i,j,k}  \quad \forall i, \forall j, \forall k \in L  \label{const:MustBeServed}\\
  &\text{Availability constraint: } x^{t+1}_{i,M+1,M+1}= 1 \quad \forall i\label{const:AllItemsInCP}.
\end{align}
Our objective (\ref{obj:MinCost}) calculates the cost of serving user requests over all the edge routers and minimizes this cost by favoring the most popular files.
Note that if $x_{i,j,j}=1$, then $f_{i}$ is stored in $R_j$.
\textit{Cache capacity constraints} in~(\ref{const:CacheCapacity}) and (\ref{const:CacheCapacityNotInS}) ensure the total size of items to be stored in a router's cache cannot exceed cache capacity.
Only routers in $\mathcal{S}$ can consider putting the requested item $f_{i}$ into their caches.
\textit{Maximum replica constraint} in~(\ref{const:Ncopy}) ensures that an item can have maximum $\Ncopy$ replicas in the network. Note that by removing this constraint, system can figure out optimal $K$ for each neighborhood automatically.
\textit{Feasibility constraint} in~(\ref{const:FetchFromOther}) reflects $f_{i}$ being retrievable from $R_{k}$ only if $R_{k}$ stores $f_{i}$ whereas~(\ref{const:NoChangeInContents}) states that contents cached by routers not in $\mathcal{S}$ do not change.
\textit{Service constraint} in ~(\ref{const:MustBeServed}) forces the content to be served from some location (i.e., local cache, another router's cache, or the CP) while \textit{availability constraint} in ~(\ref{const:AllItemsInCP}) ensures that all items are available from the CP.
Decision variables are binary, i.e., $x_{i,j,k}\in \{0,1\}$.
$C_{OPT}$ is an integer linear programming problem which can be solved with optimization software for small instances of the problem but it requires low-complexity distributed schemes for large scale networks.
We leave distributed solutions for future work.

 Let $\mathcal{F}_j = \{u_{j,1}, u_{j,2}, u_{j,3} ... \}$ be the list of user requests arriving at leaf router $R_j$ where $u_{j,i}$ is the $i^\textrm{th}$ request for a file with size $s_{u_{j,i}}$. $R_j$ can retrieve it only from its reachable set $\ReachableSet_j$ which is defined as
 
\begin{align}    \ReachableSet_j=\bigcup_{R_k\in\mathcal{R}_{j,CP}}\SearchableSet_k\label{eq:ServingRouterSetDefinition}.
\end{align}
 
 A request will be counted as hit if at least one of the routers in $\ReachableSet_j$ stores it. More formally, we define hit function $\delta_{j,i}$ for request $u_{j,i}$ (assuming $u_{j,i}$ is a request for $f_i$) as follows:
     $$
      \delta_{j,i} = \begin{cases} 1 &\mbox{if } \sum_{R_k\in \ReachableSet_j} x_{i,k,k} \geq 1 \\
      0 & \mbox{ o/w.}  \end{cases}
     $$ Next, we calculate BHR as follows:
    \begin{align}
    BHR &=\frac{\sum_{j=1}^{L}{\sum_{\forall u_{j,i}\in\mathcal{F}_j}s_{u_{j,i}} \delta_{j,i}}}{\sum_{j=1}^{L}{\sum_{\forall u_{j,i}\in {\mathcal{F}_j}}s_{u_{j,i}}}}. \label{eq:BHR}\end{align}
    If request $u_{j,i}$ is served from a router that is $h_{j,i}$ hops away from the user and the path from $R_j$ to the CP is $H_j$ hops long, we can compute the FPR as follows:
    \begin{align}
      FPR &= 1 - \frac{\sum_{j=1}^{L}{\sum_{\forall u_{j,i}\in\mathcal{F}_j}s_{u_{j,i}} h_{j,i}}}{\sum_{j=1}^{L}H_j{\sum_{\forall u_{j,i}\in {\mathcal{F}_j}}s_{u_{j,i}}}}. \label{eq:FPR}
     \end{align}

\section{Numerical Analysis}
\label{sec:evaluate}

\subsection{Setup \& Metrics}
\label{sec:evaluate:setup}

We performed numerical evaluation on realistic and synthetic topologies.
Realistic topologies are from~\cite{SpringN:Rocketfuel}, and synthetic topologies are scale-free networks of 50 nodes.
Each node can store 25 objects.
We present results on synthetic networks; realistic topologies produce similar results.
Content popularity is modeled according to~\cite{Cha:2007:ITY:1298306.1298309}, and content set contains 5000 objects.
We calculate the betweenness centrality ($C_B$) of each router in order to analyze its impact on cached content in a specific router under various $(K, r)$ pairs.
We define \emph{coupling factor} (\CouplingCorrelation) as the Pearson correlation between $C_B$ and average popularity per bit in a node's cache; it measures topological impact on system performance.
The rationale is that optimal system performance is achieved by placing content at specific locations in a network according to its popularity and that $C_B$ is a good metric to characterize a node's position in a graph.
Strong correlation between the two indicates that content is tightly ``coupled'' with topology and topological properties influence system performance.

 In the simulations, 30\% of the edge routers are randomly selected and connected with client, and the server randomly connects to one of the 5 core nodes with highest $C_B$.
Experiments were repeated at least 50 times.

\subsection{Pareto Frontier}
\label{sec:evaluate:pareto}

% \begin{figure*}[!htp]
%   \centering \subfloat[Pareto frontier of in-network caching
%   performance.]
%   {\label{fig:eval:1:pareto}\includegraphics[width=4.2cm]{figure/pareto_frontier}}
%   \quad 
%   \subfloat[Results as function of $\Ncopy$ with $\radius=0$. Segment $\overline{AC}$ in Fig.
%   \ref{fig:eval:1:pareto}]{\label{fig:eval:1:1}\includegraphics[width=4.2cm]{figure/fig_pareto_ac}}
%   \quad 
%   \subfloat[Results as function of $\radius$ with $\Ncopy=1$. Segment $\overline{AB}$ in Fig.  \ref{fig:eval:1:pareto}]
%   {\label{fig:eval:1:2}\includegraphics[width=4.2cm]{figure/fig_pareto_ab}}
%   \quad 
%   \subfloat[Results as function of $\Ncopy$; moving from B to C on Pareto
%   frontier.]{\label{fig:eval:1:3}\includegraphics[width=4.2cm]{figure/fig_pareto_bc}}
%   \caption{Cooperation policy pushes the system performance to its
%     Pareto frontier by the interplay of $\radius$ and $\Ncopy$.}
%   \label{fig:eval:1}
%   \vskip -3mm
% \end{figure*}

\begin{figure}[!tb]
  \centering 
  \includegraphics[width=8cm]{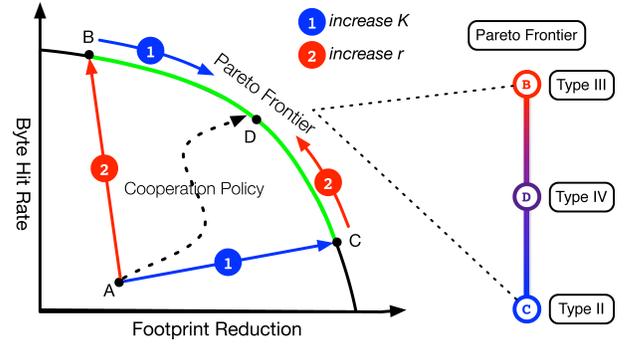} 
  \caption{Conflicting BHR and FPR at the Pareto frontier.
  Type of cooperation at different points shown on right.}
  \label{fig:eval:1:pareto}
\end{figure}

%\begin{figure*}[!tb]
%  \centering
%\includegraphics[width=15cm]{figure/fig_pareto_combined}
%  \caption{Change in BHR (Top plot, left y-axis), FPR (top, right y-axis), and CPF (bottom) along $AB$, $BC$, and $CA$.}
%\label{fig:eval:1:1}%  \label{fig:eval:1}
%  \vskip -3mm
%\end{figure*}
%

\begin{figure*}[!tb]
  \centering
  \quad \subfloat[Change in BHR (Top plot, left y-axis), FPR (top, right y-axis), and CPF (bottom) along $AB$, $BC$, and $CA$.]{\label{fig:eval:1:1}\includegraphics[width=12.5cm]{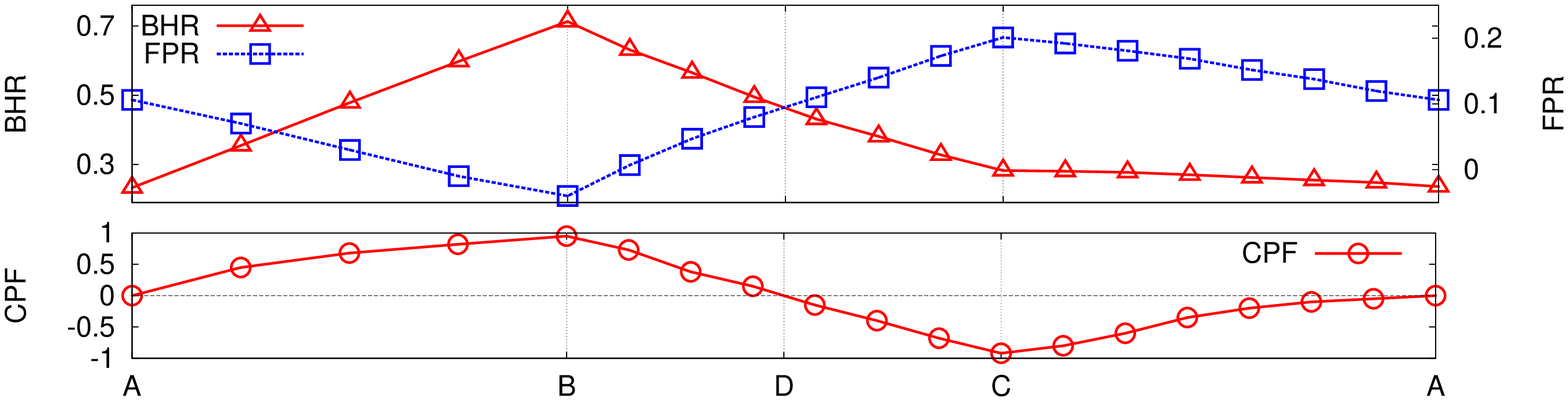}}
  \quad \subfloat[CPF, BHR, and FPR as a function of $K$ and $r$.]{\label{fig:eval:3:1}\includegraphics[width=3.3cm]{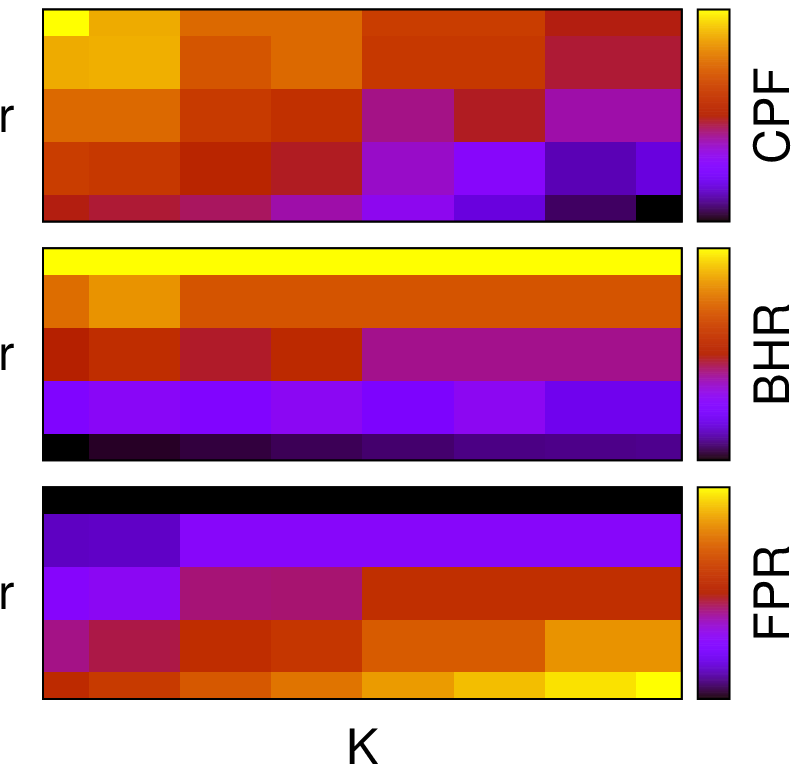}}
  \caption{Performance of $(K,r)$-Cooperation along the boundary defined by ABC (a), and for all $(K,r)$ pairs (b).}
\label{fig:eval:1:0}%  \label{fig:eval:1}
\end{figure*}

Fig.~\ref{fig:eval:1:pareto} shows how $\Ncopy$ and $\radius$ impact caching performance.
The solution to $C_\mathrm{OPT}$ provides the optimal cache profiles for given $\Ncopy$ and $\radius$ (e.g., point $A$ in Fig.~\ref{fig:eval:1:pareto}), but it does not indicate the best values for these two parameters, i.e., we can improve performance by tuning $\Ncopy$ and $\radius$, because the system may be underutilized.
However, our optimization model can be used to find Pareto frontier of the performance (green arc $BC$ in Fig.~\ref{fig:eval:1:pareto}).
When we reach the Pareto frontier, we cannot improve BHR or FPR without hurting the other.
The fan-shaped area defined by $ABC$ is the area which a cooperation policy can explore to find the best tradeoff between $\Ncopy$ and $\radius$.
Point $D$ where we eventually reach the Pareto frontier depends on how cooperation policy balances BHR and FPR.
Lines $AB$ and $AC$ are not parallel to the x- and y-axis, since changing either of $\radius$ or $\Ncopy$ affects both BHR and FPR, as we show below.

%\textbf{NEED NEW VERSIONS OF FIGURES FOR THIS PART}

The upper graph in Fig.~\ref{fig:eval:1:1} shows how BHR and FPR vary as we move along the segments $AB$, $BC$, and $CA$, by varying $\radius$ and $\Ncopy$.
%how performance changes along the segments $AB$, $AC$ and arc $BC$ by increasing $\radius$ and $\Ncopy$ respectively.
Starting from $A$ and moving clockwise (left to right in the figure), we increase the search radius which improves BHR, but decreases FPR due to additional search traffic or letting content be cached at routers with higher $h_{j,i}$ in (\ref{eq:FPR}).
From $B$ to $C$, along the Pareto frontier, we observe the tradeoff between BHR and FPR, with FPR reaching its maximum at $C$.
From $C$ to $A$, $\radius$ is 0 so the system reduces to en-route caching where larger number of copies (near $C$) is beneficial, hence as we move towards $A$, both BHR and FPR decrease. Fig.~\ref{fig:eval:3:1} shows heatmaps of CPF, BHR, and FPR as function of $\Ncopy$ and $\radius$.
Lighter values indicate higher values.
It shows how BHR and FPR conflict each other, i.e., one achieving the highest performance while the other has the worst, in regions corresponding to the Pareto frontier.

\begin{figure*}[!tb]
  \centering
  \vskip +5mm
  \quad \subfloat[Point $B$ in
  Fig.~\ref{fig:eval:1:pareto}.]{\label{fig:eval:3:2}\includegraphics[width=4.3cm]{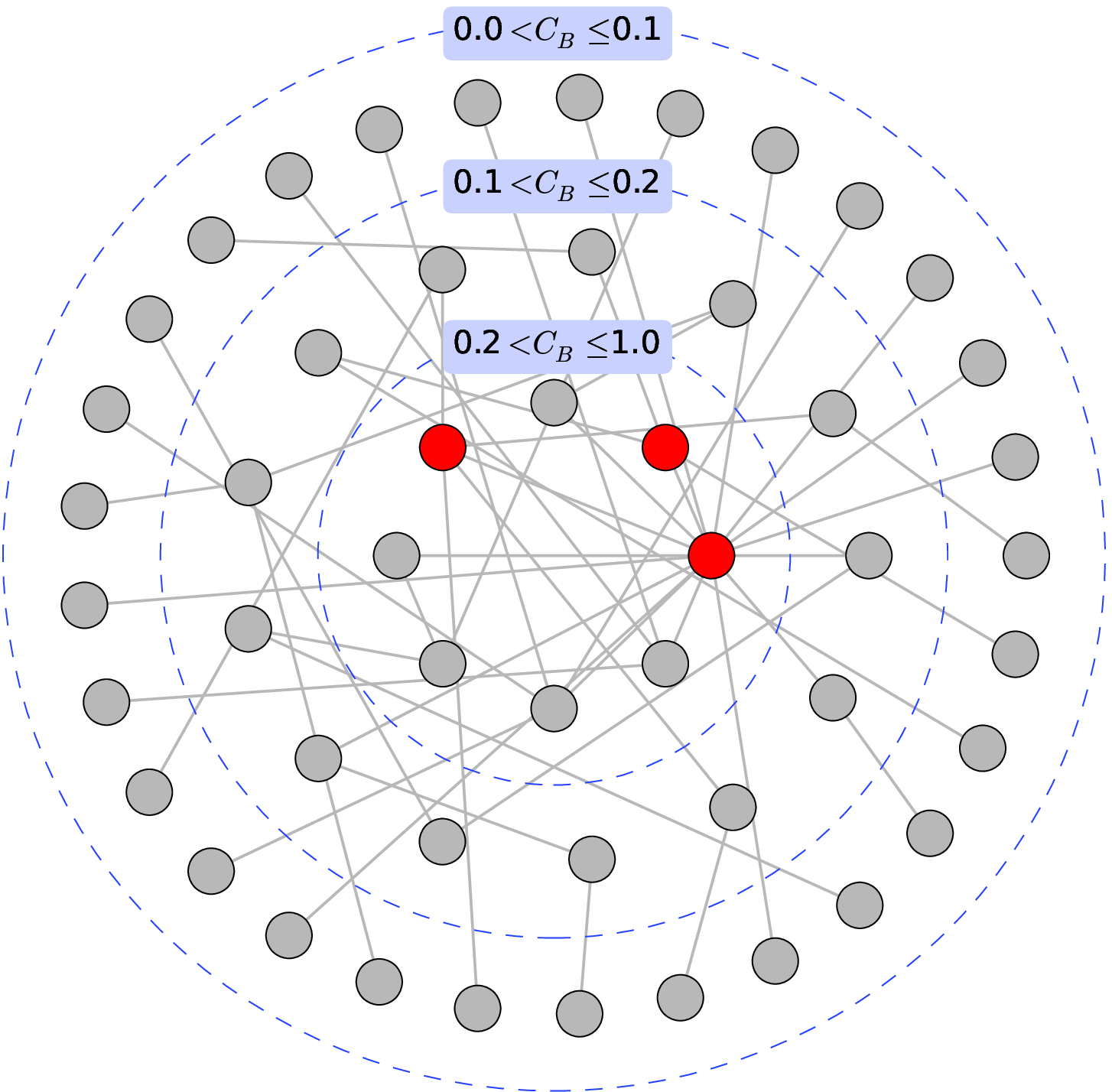}}
  \quad \quad \quad \subfloat[Point $D$ in
  Fig.~\ref{fig:eval:1:pareto}.]{\label{fig:eval:3:3}\includegraphics[width=4.3cm]{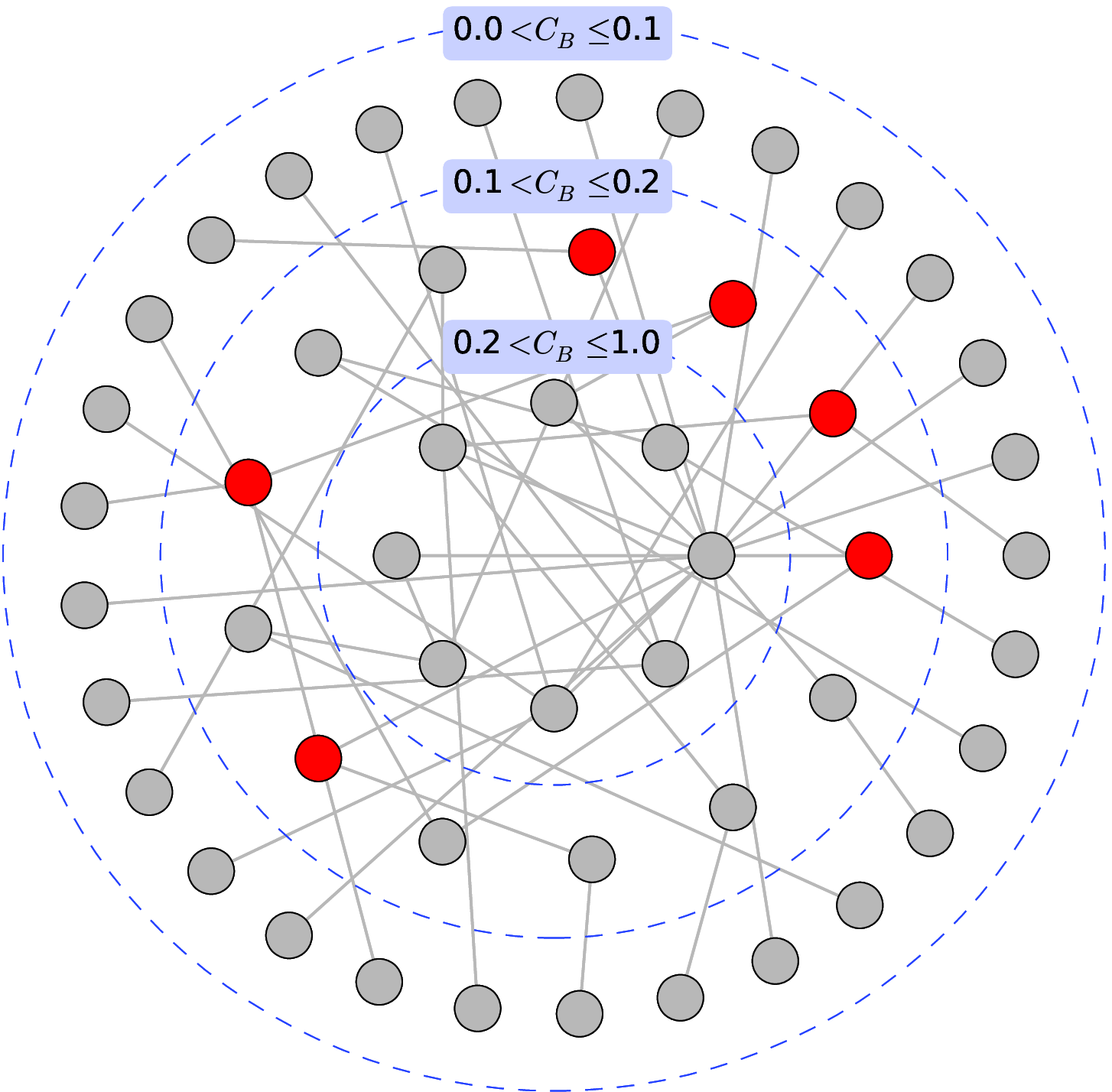}}
  \quad \quad \quad \subfloat[Point $C$ in
  Fig.~\ref{fig:eval:1:pareto}.]{\label{fig:eval:3:4}\includegraphics[width=4.3cm]{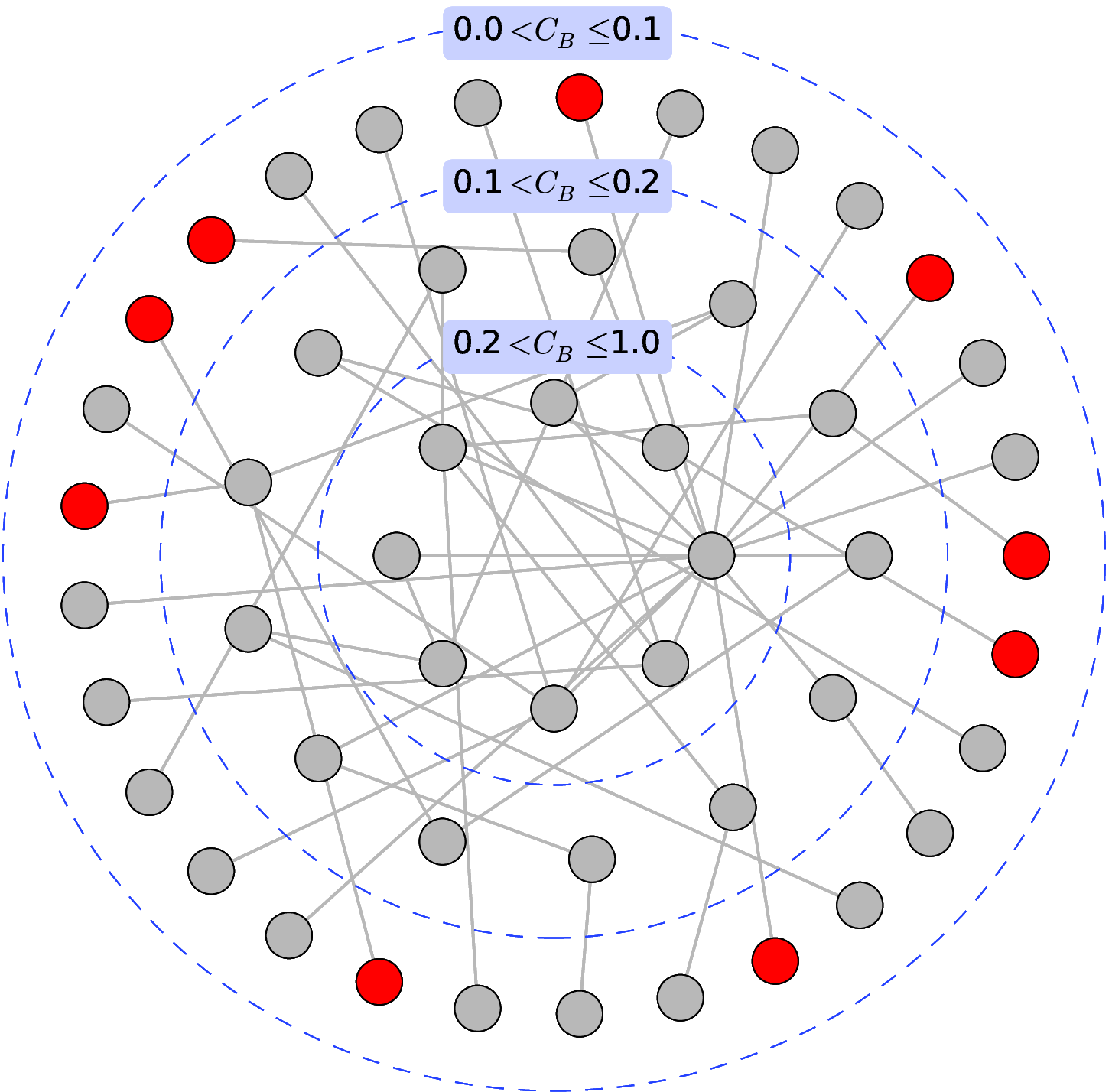}}
  \caption{Content placement at points $B$, $D$, and $C$.
    Red color (dark dots) marks the most popular content.
    Nodes are grouped in circles according to $C_B$.
    Content migrates from core to edge as we move from $B$ to $C$.
% Change in popular content placement from point B to D and to C. Red colour marks the top 5\% of the most popular content,
%     nodes are grouped in concentric circles according to their
%     betweenness centrality. Cooperation policy is controlled by
%     $\Ncopy$ and $\radius$. Figure shows how cooperation policy
%     impacts the content distribution in the network.
  }
  \label{fig:eval:3}
  \vskip -5mm
\end{figure*}

\subsection{Coupling Content and Topology}
\label{sec:evaluate:isp}

% How to take advantage of network topological properties to enhance caching performance is of great interest in ICN research.
% If successfully utilized, such properties should have significant impacts on performance.
% We promote the idea that of coupling the content with topology and argue that it makes sense to utilize those properties (e.g.
% $C_B$) only when tight coupling exists.

%We now show the relationship between cooperation policy, content popularity, and network topological properties.

The lower plot in Fig.~\ref{fig:eval:1:1} shows how CPF evolves along the same path.
%Fig.~\ref{fig:eval:1:1} and \ref{fig:eval:1:2} show $\CouplingCorrelation$ as a function of one of the two cooperation parameters given the other is fixed.
Values close to -1 or 1 indicate strong dependence between popularity and betweenness.
%The result is very interesting in terms of how cooperation policy couples the content popularity with topological properties.
A router with a high $C_B$ is in or close to the core of the network whereas a router with low $C_B$ is close to the network edge.
%\textbf{REVISE AFTER NEW FIGURE INSERTED}
At $B$, where CPF is close to 1, popular content is in nodes with high $C_B$, i.e., the core, whereas at $C$, where CPF is close to -1, it is at the edge where $C_B$ is low.
Along the Pareto frontier $BC$, we observe a ``migration'' of content from core to edge.
At $D$ where CPF is 0, both BHR and FPR are close to halfway point between their respective minima and maxima at $B$ and $C$.
We have observed this phenomenon across a wide range of experimental settings, but its full investigation is left for further study.

% With this in mind, Fig.~\ref{fig:eval:3:1} shows that given $r$, increasing $K$ will decrease the correlation to almost $-1$, which means the popular content is pushed to the network edge (reflected in Fig.~\ref{fig:eval:3:4}).
% On the contrary, given $K$, increasing $r$ will increase correlation to $+1$, which means the popular content is pulled to the network core (reflected in Fig.~\ref{fig:eval:3:2}).

% Figure is higher up

Fig.~\ref{fig:eval:3} shows cooperation policy's impact on content place\-ment along the Pareto frontier $BC$.
Point $B$ (Fig.~\ref{fig:eval:3:2}), representing Type~III cooperation favors BHR and places content in the core.
Point $D$ along $BC$ (Fig.~\ref{fig:eval:3:3}) strikes a tradeoff between BHR and FPR and the content is neither in the core nor on the edge; this is Type~IV cooperation.
Finally, point $C$ (Fig.~\ref{fig:eval:3:4}) favors FPR and pushes popular content to the edge.

% Figs.~\ref{fig:eval:3:2},~\ref{fig:eval:3:3}, and~\ref{fig:eval:3:4} visually illustrate how popular content migrates from the network core to the edge under the different cooperation types, which have different preferences for BHR and FPR.
% Fig~.\ref{fig:eval:3:2} corresponds to point $B$ in Fig.~\ref{fig:eval:1:pareto} and represents Type III cooperation favoring BHR.
% The most popular content gathers at the core and $\CouplingCorrelation$ is close to $1$.
% Similarly in Fig.~\ref{fig:eval:3:4}, Type II cooperation prefers FPR and pushes the popular content to the edge, driving $\CouplingCorrelation$ close to $-1$.
% Whereas in Fig.~\ref{fig:eval:3:3}, Type IV cooperation tries to be impartial between the two metrics and the content is placed neither in the core nor at the edge.

\subsection{Implications}
\label{sec:implication}

Our results have profound implications on relationship between cooperation policies, content popularity, and network topology.
They also give us hints on when and how topological properties should be taken into account in caching strategy design.
We summarize the main implications as follows:

\begin{enumerate}

\item Cooperation policy pushes performance to Pareto frontier and couples content popularity and topological properties together.
How and where it falls on the frontier depends on how it balances BHR and FPR.

\item Content popularity and topology strongly correlate with each other only close to the Pareto frontier.
Whether the correlation is positive or negative depends on how the cooperation policy favors one of the two metrics.  
%If the cooperation policy has a preference for one of the two metrics, which further determines whether it is positive or negative correlation.

\item The optimization model implies that $C_B$ has more influence on performance when we get closer to points $A$ or $B$ in Fig.~\ref{fig:eval:1:pareto}; only Type II and III cooperation policies can fully utilize $C_B$ to enhance performance.

\item We only present results on betweenness centrality.
We also experimented with other centrality measures and got similar results.
The impact of other properties like diameter or clustering coefficient needs further study.

\item We conjecture that tight coupling between content popularity and topology comes similar mathematical structures as both exhibit power-law properties.
Were popularity closer to uniform or topology closer to a random network, this tight coupling might disappear.
However, this matter requires further study. %

%  One interesting conjecture is the reason for the tight coupling between content and topology might come from the fact they have similar mathematical structure.
% In other words, Zipf popularity matters, and scale-free network matters.
% If we change one of them to uniform popularity or random network, the tight coupling may disappear.
% Whether it still makes sense trying to take advantage of topological information in such cases remains an open question.

\end{enumerate}

% \subsection{When Topology Plays a Role?}
% Experiments are needed.

% \subsection{Content Matters!}
% Experiments are needed.
 
%\begin{figure}[!tb]
%  \centering
%  \includegraphics[width=5.5cm]{figure/fig_heatmap_combined}
%  \caption{CPF, BHR, and FPR as a function of $K$ and $r$.}
% \label{fig:eval:3:1}
% \vskip -5mm
%\end{figure}
%

\section{Conclusion}
\label{sec:conclusion}

We modeled cache cooperation by its search radius and tolerance of duplicates.
We performed a thorough numerical analysis and showed that cooperation policy pushes system performance to its Pareto frontier, and how it couples content with topology.
We proposed a way to measure impact of topology on system performance.
We show when and how topological information should be taken into account in in-network caching strategy design.

%\input{text/introduction}
%\input{text/model}
%\input{text/optmodel}
%\input{text/evaluation}
% \input{text/related}
%\input{text/conclusion}

%\bibliographystyle{IEEEtran}
%\bibliography{pareto}


\end{document}